\documentclass[journal,11pt,onecolumn,draftclsnofoot,]{IEEEtran}

\usepackage{amsmath,amssymb,latexsym,float}
\usepackage{bm}
\usepackage{graphicx}
\usepackage{subfigure}
\usepackage{caption}
\usepackage{epsfig}
\usepackage{epstopdf}
\usepackage{multirow}
\usepackage{color}
\usepackage{url}
\usepackage{color}
\usepackage{array}
\usepackage{tabularx}
\allowdisplaybreaks
\usepackage{lineno}
\usepackage[colorlinks,allcolors=blue]{hyperref}
\usepackage[a4paper, total={7in, 9in}]{geometry}
\newcommand{\R}{\mathbb{R}}
\newcommand{\C}{\mathbb{C}}
\newcommand{\F}{\operatorname{F}}

\newcommand{\vct}[1]{\boldsymbol{#1}}
\newcommand{\mtx}[1]{\boldsymbol{#1}}
\newcommand{\set}[1]{\mathcal{#1}}
\newcommand{\va}{\vct{a}}

\newcommand{\vf}{\vct{f}}

\newcommand{\vx}{\vct{x}}

\newcommand{\vtheta}{\vct{\theta}}

\newcommand{\mA}{\mtx{A}}
\newcommand{\mB}{\mtx{B}}
\newcommand{\mC}{\mtx{C}}
\newcommand{\mD}{\mtx{D}}
\newcommand{\mE}{\mtx{E}}
\newcommand{\mF}{\mtx{F}}
\newcommand{\mG}{\mtx{G}}
\newcommand{\mH}{\mtx{H}}
\newcommand{\mI}{\mtx{I}}

\newcommand{\mK}{\mtx{K}}
\newcommand{\mL}{\mtx{L}}

\newcommand{\mP}{\mtx{P}}
\newcommand{\mQ}{\mtx{Q}}
\newcommand{\mR}{\mtx{R}}
\newcommand{\mS}{\mtx{S}}
\newcommand{\mT}{\mtx{T}}
\newcommand{\mU}{\mtx{U}}
\newcommand{\mV}{\mtx{V}}

\newcommand{\mX}{\mtx{X}}
\newcommand{\mY}{\mtx{Y}}
\newcommand{\mZ}{\mtx{Z}}

\newcommand{\mSigma}{\mtx{\Sigma}}

\newcommand{\setB}{\set{B}}

\newcommand{\setO}{\set{O}}

\newcommand{\setW}{\set{W}}

\newtheorem{thm}{Theorem}

\begin{document}
\title{Sub-Nyquist Sampling of Sparse and Correlated Signals in Array Processing}
\author{Ali Ahmed, Fahad Shamshad and Humera Hameed \thanks{Authors are with Information Technology University of the Punjab, Lahore, Pakistan.}}
\maketitle
\begin{abstract}
This paper considers efficient sampling of simultaneously sparse and correlated (S\&C) signals for automotive radar application. We propose an implementable sampling architecture for the acquisition of S\&C at a sub-Nyquist rate. We prove a sampling theorem showing exact and stable reconstruction of the acquired signals even when the sampling rate is smaller than the Nyquist rate by orders of magnitude. Quantitatively, our results state that an ensemble $M$ signals, composed of a-priori unknown latent $R$ signals, each bandlimited to $W/2$ but only $S$-sparse in the Fourier domain, can be reconstructed exactly from compressive sampling only at a rate $RS\log^{\alpha} W$ samples per second. When $R \ll M$ and $S\ll W$, this amounts to a significant reduction in sampling rate compared to the Nyquist rate of $MW$ samples per second. This is the first result that presents an implementable sampling architecture and a sampling theorem for the compressive acquisition of S\&C signals. We resort to a two-step algorithm to recover sparse and low-rank (S\&L) matrix from a near optimal number of measurements. This result then translates into a signal reconstruction algorithm from a sub-Nyquist sampling rate. 
	
%	The signal reconstruction from sub-Nyquist rate boils down to a sparse and low-rank (S\&L) matrix recovery from a few linear measurements. The conventional convex penalties for S\&L matrices are provably not optimal in the number of measurements. We resort to a two-step algorithm to recover S\&L matrix from a near optimal number of measurements. This result then translates into a signal reconstruction algorithm from a sub-Nyquist sampling rate. 

\end{abstract}



%\begin{keyword}
%sparse \sep correlation \sep low-rank \sep compressive sensing
%\end{keyword}

%\end{frontmatter}

\section{Introduction}\label{sec:intro}

Automotive radar (AR) plays an indispensable role in the development of autonomous vehicles and advanced driver assistance systems (ADASs) \cite{pole2017automotiveat}. Digital computation is deeply ingrained in modern signal processing algorithms behind ARs, and an efficient analog-to-digital conversion is of fundamental importance.  
This paper proposes a novel sampling architecture for the acquisition of a simultaneously sparse and correlated (S\&C) signal ensemble at a sub-Nyquist rate.  An S\&C ensemble consists of multiple signals well-approximated by the linear combinations of a few latent signals that are also sparse in some transform domain. Such ensembles arise in various applications in array processing \cite{ahmed2015compressive2,ahmed2015compressive}, where it is easy to come across thousands of signals possibly spanning wide bandwidths  \cite{sridharan1998us,maskell2008sapphire,larsson2014massive} but with a lot of latent redundancies that can be well-approximated using S\&C structure. %For example, in neurophysiology micro electrode arrays with thousands of recording sites are employed to study the neural activity in the brain tissue \cite{haas08pr,gray04di}. Generally, a neuron fires at most a few times every second leading to a sparse electrical signal. Moreover, very often the activity at a given time is limited to a few neurons and the remaining are dormant. This leads to a multiple array elements recording similar or very correlated signals often. 
\begin{figure}
	\centering
	 \includegraphics[trim = {0cm 6cm 8cm 0cm}, scale = 0.6]{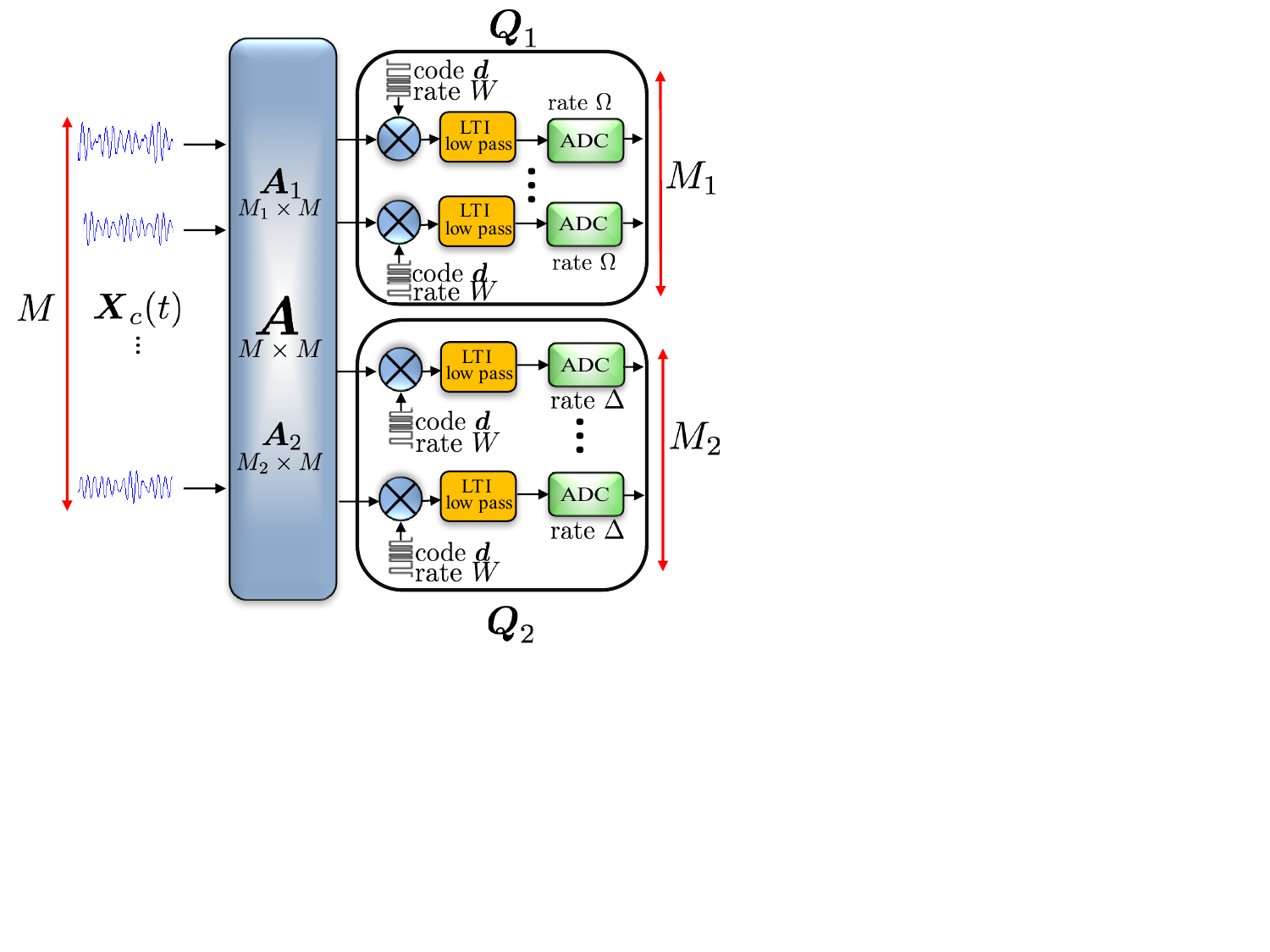}
	\caption{\small \sl Sampling Architecture: $M$ signals in the ensemble $\mX_c(t)$ are mixed across channels using an analog-vector-matrix multiplier (AVMM) and then modulated (multiplication by a random binary waveform), low-pass filtered (using an integrator), and eventually sampled at a rate $\Omega$ in top $M_1$ branches and at rate $\Delta$ in the remaining bottom $M_2$ branches ($M_1+M_2 = M$).  We show that when the total sampling rate  $M_2\Delta + M_1\Omega$ roughly exceeds $RS\log^{\alpha} W$ samples per second --- a significant improvement over the rate $MW$ dictated by Shannon-Nyquist sampling theorem, enables stable signal reconstruction.}
	\label{fig:Sampling-Architecture}
\end{figure} 
Acquiring such an ensemble of large number of signals plainly at the Nyquist rate in some applications including ADAS produces data on the order of several gigabits to terabits per second. Transferring such a humongous amount of data off-chip becomes a significant challenge, especially, for prolonged monitoring. In addition, the cost of an analog-to-digital converter (ADC) ramps up rapidly with increasing sampling rates, and the precision (quantization levels) of the collected samples also decreases with faster sampling rates. Moreover, for several on-chip applications, the power dissipation needs to be controlled, and a faster ADC always requires more power and leads to a larger dissipation. An on-the-fly, sub-Nyquist rate acquisition of such a spatially and temporally redundant signal ensemble is, therefore, of practical significance especially in modern ARs. It is important to note that the sub-Nyquist sampling is a challenging proposition as the signal sparsity, and correlation pattern among the signals is not known a priori, and hence cannot be leveraged to collect fewer, and strategically placed non-redundant spatial and temporal samples to design a sub-Nyquist sampling scheme. 

Using Shannon-Nyquist sampling theorem, an ensemble of $M$ signals, each bandlimited to $W/2$ Hz can be acquired at $MW$ uniform samples per second. We show that if every signal in the ensemble is a superposition of underlying fewer number $R$ of signals (correlated) that have only $S$ active frequency components (sparse) then the ensemble can be acquired by sampling only at a much lower rate of roughly $RS$ samples per second, which is indeed a significant reduction of the sampling rate, especially when $R \ll M$, and $S \ll W$.  We design a sampling architecture; shown in Figure \ref{fig:Sampling-Architecture}, using simple-and-easy-to-implement components such as switches, and integrators for the preprocessing of analog signals. Each signal is then compressively sampled using a low-rate ADC. 
\begin{figure}
	\centering
	\includegraphics[scale = 0.42, trim = {0cm 0cm 0cm 0cm}]{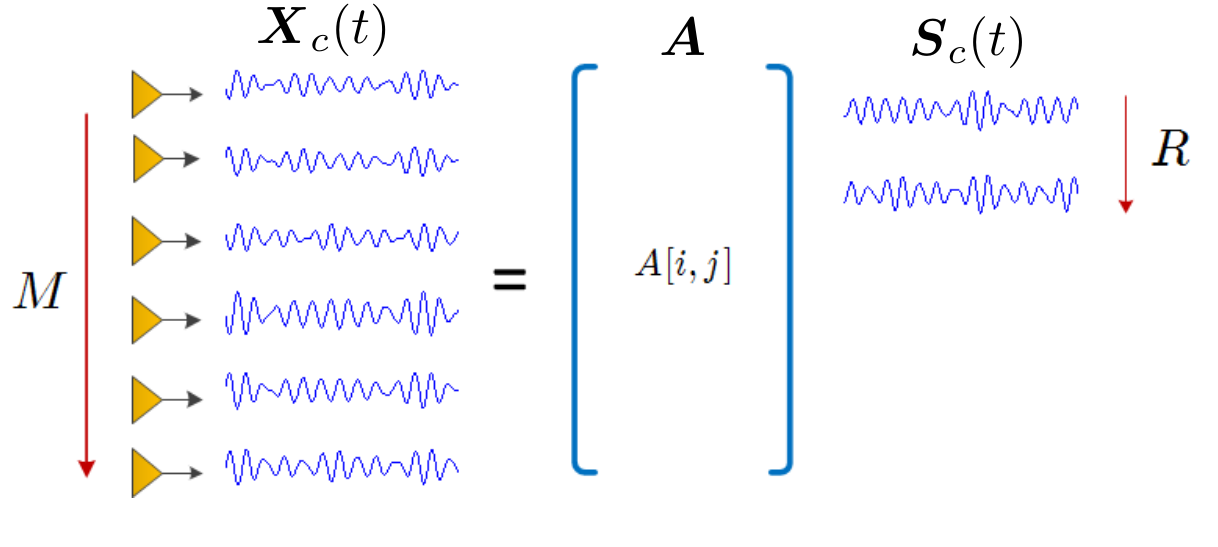}
	\caption{\small \sl Spectrally Sparse and Correlated (S\&C) Signal Ensemble: Signal Ensemble $\mX_c(t)$ is composed of $M$ signals, each bandlimited to $W/2$. Each of the signal in the ensemble is a superposition of underlying fewer $R$ signals in the ensemble $\mS_c(t)$. The signals in the ensemble $\mS_c(t)$ contain only $S$ unique active frequency components.}
		\label{fig:Prob-intro}
\end{figure}

Compressive sampling of spectrally sparse signals has been a topic of interest in recent years \cite{tropp2010beyond,laska07th, Ding2020CS}. The signal reconstruction from a few samples is framed as a sparse-recovery problem from a limited number of measurements and is handled efficiently using an $\ell_1$ minimization program.
Similarly, compressive sampling of correlated signals is studied in  \cite{ahmed2015compressive,ahmed2015compressive2,ahmed2017compressive,ahmed2011compressive,ahmed2012compressive,ahmed2013compressive}. In this case, the reconstruction of signal ensemble from a few samples is recast as a low-rank matrix recovery problem from a limited number of measurements, which is effectively solved using a nuclear-norm minimization program. In this paper, we show that compressive sampling of a simultaneously sparse and correlated signal ensemble boils down to recovering a simultaneously sparse and low-rank (S\&L) matrix from a few linear measurements. A natural choice of solving an $\ell_1$ plus nuclear-norm minimization program, however, does not lead to S\&L matrix recovery from an optimal number of measurements \cite{oymak2015simultaneously}. This problem obstructs the acquisition of S\&C signal using low-rate ADCs. 

\textit{\textbf{Our Contributions:}} In this paper, we overcome this problem with a new signal reconstruction algorithm consisting of two steps: $\ell_1$ minimization followed by a least-squares program, to recover the S\&L matrix from a near optimally few numbers of measurements. This result directly translates into S\&C signal reconstruction at a sub-Nyquist rate. Specifically, we design an implementable sampling architecture to acquire an S\&C signal ensemble at potentially well-below the Nyquist sampling rate, and a computationally efficient, and novel algorithm to recover the signal ensemble from the acquired compressive samples. We rigorously prove that the proposed algorithm can recover the S\&C ensemble from an optimally fewer compressive samples, and give a formal statement of this result as a sampling theorem. 

\textit{\textbf{Organization of the paper:}} We start by introducing the signal structure more precisely in Section \ref{sec:Signal-Model}. We briefly comment on the implementation aspect of the proposed sampling architecture in Section \ref{sec:Sampling-Architecture}. The samples collected using the ADCs are expressed as a linear transformation of the input signal ensemble in Section \ref{sec:DiscreteForm}. Section \ref{sec:Column-Row-Measurements} and  \ref{sec:Algo} present the signal reconstruction algorithm with numerical simulations in Section \ref{sec:Numerics}. A summary of the notations used in this paper is presented in Table [1] for convenience.

\begin{table}[t] 
\label{table:notations}
\centering
\scalebox{0.9}{
\begin{tabular}{|l|l|}
\hline
\textbf{Notation}     & \multicolumn{1}{c|}{\textbf{Description}} \\ 
\hline
$\mX_{c}(t)$  &  A matrix with continuous-time correlated and sparse signals,\\ 
& $\{\vx_1(t), \ldots, \vx_M(t)\}$ as its rows\\\hline
$\boldsymbol{x}_m(t)$  & $m$th row of  $\mX_{c}(t)$.\\ \hline
$C[m,\omega]$ & DFT coefficient of $\vx_m(t)$  at frequency $\omega$\\ \hline
$\mC$ & An $M \times W$ matrix with $C[m,\omega]$ as the $(m,\omega)$th entry.\\ \hline
$\Gamma_m$ & Support of non-zero frequencies in the Fourier spectrum of $\vx_m(t)$\\ \hline
$M$ & Number of signals in the ensemble\\ \hline
$S$ &  Upper bound on $|\Gamma_1 \cup\ldots\cup\Gamma_M|$\\ \hline
$R$  & Rank of signal ensemble  $\mX-c(t)$\\ \hline
$B$ & Maximum bandwidth (Hz) of the signals in the ensemble\\ \hline
$\mX$ & An $M \times W$ matrix of samples of $\mX_c(t)$, \\
& where $W = 2B+1$, and $\mX = \mC\mF^*$ \\ \hline
$\mF$ & $W \times W$ normalized DFT matrix \\ \hline
 $\mA$  &  An  $M \times M$ random orthonormal mixing matrix\\ \hline
$\mA_1$& Top $M_1$ rows of $\mA$, $M_2 = M-M_1$\\ \hline
$\mA_2$& Bottom $M_2$ rows of $\mA$\\ \hline
$\Omega$  & Sampling rate for top $M_1$ signals in $\mX_c(t)$\\ \hline
$\Delta$ & Sampling rate for the bottom $M_2$ signals in $\mX_c(t)$ \\ \hline
$b(t)$  & Random binary $\pm 1$ waveform \\ \hline
$\mD$ & Diagonal matrix $\text{diag}(b[1],\ldots,b[W])$ \\ \hline
$\mT$ & $W \times W$ diagonal matrix with entries  $T[\omega,\omega] = \tfrac{\mathrm{e}^{\iota 2\pi \omega/W}-1}{\iota 2\pi \omega}$ \\ \hline
$\mH$ & Unknown $M \times W$ matrix. $\mH = \mC\mT$\\ \hline
$\mP_{\Omega, W}$ & $\Omega \times W$ matrix. $\mP_{\Omega,W}\vx$ returns length\\& $\Omega$ vector by summing consecutive $W$/$\Omega$ entries of $\vx$ \\ \hline
$\mP_{\Delta, W}$ & $\Delta \times W$ matrix. $\mP_{\Delta,W}\vx$ returns length\\& $\Delta$ vector by summing consecutive $W$/$\Delta$ entries of $\vx$ \\ \hline
$\mQ_1$ & $\mF_{\Omega, W}\mD\mF$ \\ \hline
$\mQ_2$  & $\mP_{\Delta, W}\mD\mF$  \\ \hline
$\mY_1$ & $\mA_1\mC\mT\mQ_1^*+\mE_1$ and $\mE_1$ is the noise matrix, where $\|\mE_1\|_{\F} \leq \delta_1$ \\ \hline
$\mY_2$ & $\mA_2\mH\mQ_2^*+\mE_2$ and $\mE_2$ is the noise matrix, where $\|\mE_2\|_{\F} \leq \delta_2$  \\ \hline
$\mH_R$ & Best rank-$R$ approximation of $\mH$\\ \hline
$\hat{\mH}$ & Estimate of $\mH$\\ \hline
\end{tabular}}
\caption{Summary of the notations used in the paper.}
\end{table}

\section{Signal Model}\label{sec:Signal-Model}
We consider an ensemble $\mX_c(t)$ of $M$ continuous-time correlated and sparse signals $x_1(t),x_2(t),\ldots, x_M(t)$. By correlated, we mean that every signal in the ensemble can be approximated by the linear combination of underlying minimum number $R$ of a priori unknown
signals $s_1(t), s_2(t), \ldots, s_R(t)$, that is, $x_m(t) \approx \sum_{r=1}^R A[m,r]s_r(t)$, where $A[m,r]$ are also unknown and are the entries of $\mA \in \R^{M\times R}$. Denote the smaller ensemble of $s_r(t)$'s to be $\mS_c(t)$. In the rest of the manuscript, we will think of $\mX_c(t)$, and $\mS_c(t)$ as matrices that contain the continuous time signals $x_m(t)$'s, and $s_r(t)$'s as their rows, respectively. This gives us the relation 
\begin{align}\label{eq:correlation-structure}
\mX_c(t) \approx \mA \mS_c(t). 
\end{align}
The correlation structure is illustrated in Figure \ref{fig:Prob-intro}.
 Every signal $x_m(t)$ is bandlimited\footnote{ To avoid clutter, we also take $x_m(t)$ to be periodic and, therefore, only need to consider recovery in a finite window of time (We take this window to be $t \in [0,1)$ without loss of generality).  However,  the results can be extended to non-periodic signals using smooth functions to avoid edge effects due to windowing; for details, see \cite{ahmed2015compressive,ahmed2015compressive2}. } to $B$, and its DFT is 
\begin{align}\label{eq:signalDFT}
& x_m(t) = \sum_{\omega \in \setW} C[m,\omega] \mathrm{e}^{-\iota 2\pi \omega t}, ~\text{where}~\\
&~~  t \in [0,1), ~ \text{and} ~ \setW := \{-B,\ldots, B\},\notag
\end{align}
where $C[m,\omega]$ is the $\omega$th Fourier coefficient of the $m$th signal $x_m(t)$, and also $C[m,-\omega] = C^*[m,\omega]$ as $x_m(t)$ are real. 
Define a support set of the non-zero Fourier coefficients of every $x_m(t)$ as $\Gamma_m := \{ \omega \in \setW ~|~ C[m,\omega] \neq 0\}.$ By sparse, we mean that the joint frequency band $\Gamma := \Gamma_1 \cup\cdots \cup \Gamma_M$ is sparsely occupied, and the number of non-zero frequencies in the joint frequency band $\Gamma \subset \setW$ are
\begin{align}\label{eq:Sparse-Structure}
|\Gamma| \leq S.
\end{align} 
The signal ensemble $\mX_c(t)$ is sparse in the sense of \eqref{eq:Sparse-Structure} and correlated in the sense of \eqref{eq:correlation-structure}. Observe that by definition, $R$ can only be as big as $S$ in the worst case. To see this, observe from \eqref{eq:Sparse-Structure} that every signal in $\mX_c(t)$ can be expressed as the linear combination of $S$ complex Fourier exponentials in the set $\{\mathrm{e}^{-\iota 2\pi \omega t/W} ~ |~ \omega \in \Gamma\}$. Since $R$ is the minimum number of underlying signals spanning the signal space, we have $R \leq S$ without loss of generality.  In other words, the correlation structure \eqref{eq:correlation-structure} is only non-redundant when $R$ is strictly smaller than $S$, as in this case the underlying signals $\mS_c(t)$ are not the conventional Fourier exponentials, and present an additional structure not captured by \eqref{eq:signalDFT} alone. We will see in Section \ref{sec:Applications} that in several applications in array processing $R$ is actually much smaller than $S$ and imposing the additional correlation structure leads to a reduction in the sampling rate that cannot be achieved by only imposing the spectral sparsity. 

Every signal $x_m(t)$, bandlimited to $B$ Hz, can be captured perfectly by taking a $W = 2B+1$ equally spaced samples per second (placed in the $m$th row of $M \times W$ matrix $\mX$)--- a total of $MW$ samples per second for all the signals in $\mX_c(t)$. Let $\mF$ be a $W\times W$ normalized DFT matrix with entries 
\begin{align}\label{eq:DFT}
 F[\omega,n] = \tfrac{1}{\sqrt{W}} \mathrm{e}^{\iota 2\pi\omega n}, ~ \omega \in \setW, \ \text{and}  \ n \in \{0,1,2,\ldots,W-1\}. 
\end{align}
We can write
\begin{equation}\label{eq:X-def}
\mX = \mC\mF^*, 
\end{equation}
where $C[m,\omega]$ in \eqref{eq:signalDFT} are the entries of $M \times W$ matrix $\mC$. Observe that $\mC$ is only rank-$R$, and at most $S$-sparse along the row vectors. The low-rank structure is inherited from the correlations in \eqref{eq:correlation-structure}, and row sparsity is derived from the sparsely occupied frequency band \eqref{eq:Sparse-Structure}. Taking both of these structures into account means that $\mC$ really only carries $RS$ degrees of freedom\footnote{The degrees of freedom in a rank-$R$ matrix with $S$-sparse rows are exactly $MR+RS-R^2$, which we will approximate by $RS$ in the manuscript; assuming a realistic case of $S \geq M$. Also note that $MR+RS-R^2$ is the number of unknowns, assuming the support of the non-zeros in the rows and a bases spanning the row, and column space of $\mC$ were known in advance.}, which is much smaller than the number $MW$ of samples prescribed by Shannon. This is especially true in the case of $R\ll M$, and $S\ll W$. Sparse and low-rank (S\&L) matrix $\mC$ is all that is to be determined for the reconstruction $\mX_c(t)$ in $t \in [0,1)$ from $\mX$ using sinc interpolation.  

\section{Sampling Architecture}\label{sec:Sampling-Architecture}
The sub-Nyquist rate acquisition is accomplished by a careful preprocessing of the signals in analog prior to sampling. The ensemble $\mX_c(t)$ is first processed by an analog-vector-matrix multiplier (AVMM) that takes the random linear combinations of $M$ input signals to produce $M$ outputs. This operation spreads signal energy across channels. Each signal is then modulated, which amounts to a pointwise multiplication of the signal with a random binary waveform, alternating at a rate $W$. Modulation disperses signal energy across frequency domain. The resultant signals are low-pass filtered (LPF), and a subset (top few) $M_1$ of $M$ output signals are sampled at a rate $\Omega <  W$, and the remaining $M_2$ signals at a rate $\Delta < W$, where $M = M_1+M_2$.

A word about the implementation aspect: The AVMM blocks with hundreds of inputs and outputs with a bandwidth of tens to hundreds of megahertz have been built in the recent past \cite{schlottmann2011highly,chawla2004531}. On the other hand very fast-rate modulators can be implemented using switching circuits. Modulators have already been employed in practically implementable architectures proposed for the compressive sampling of a different structured class of signals; namely, spectrally sparse signals; detail can be found in \cite{tropp2010beyond, mishali2009blind} along with the discussions on the implementation aspects of the modulators.  Low-pass filters can be easily implemented using integrators. 
\section{Observations in Matrix Form}\label{sec:DiscreteForm}
In this section, we present the discrete time formulation of the action of each of the architectural components on the input ensemble $\mX_c(t)$. We use these models to express the compressive samples acquired by the ADCs as a linear transformation of the unknown S\&L matrix $\mC$. 

Analog-vector-matrix multiplier mixes the signals by taking random linear combinations of $M$ input signals to produce $M$ outputs. Mathematically, the outputs of the AVMM are $\mA\mX_c(t)$, where we pick $\mA$ to be an $M \times M$ random orthogonal matrix:
\begin{align}\label{eq:ATA}
\mA^*\mA = \mI.
\end{align}
We denote the signals in $\mA\mX_c(t)$ by $\tilde{x}_1(t), \ldots, \tilde{x}_M(t)$. Since mixing is a linear operation, the matrix of Fourier coefficients of $\mA\mX_c(t)$ is 
\begin{align}\label{eq:tildeC}
\widetilde{\mC} : = \mA\mC, 
\end{align}
where $\mC$ is defined in \eqref{eq:X-def}. Modulator simply takes the analog signals $x_m(t)$ and returns the pointwise multiplication $x_m(t)b(t)$.  We will take $b(t)$ to be a random binary $\pm 1$ waveform that is constant $b(t) = b[k]$ over a time interval $t \in [\tfrac{k-1}{W}, \tfrac{k}{W})$, where $b[k] = \pm 1$ with equal probability. The sign changes of the binary waveforms in each of these intervals occur randomly, and independently. In other words, a modulator only shifts signal polarity from instant to instant. This will disperse the spectrum of the signals across the entire band $\setW$. Modulator in every channel uses the same binary waveform.  An $\Omega$-LPF-ADC block operates by integrating a signal over an interval $ t \in [\tfrac{(n-1)}{\Omega}, \tfrac{n}{\Omega}),\ n \in  [\Omega]$, where, in general, we define the notation $[\Omega] := \{1,2,3,\ldots,\Omega\}$. The resulting piecewise constant signal is sampled at a rate $\Omega$. In an exactly similar manner, we can also define  $\Delta$-LPF-ADC block. 

In the sampling architecture, $M$ signals at the output of the modulators are split into $M_1$ signals each of which is sampled using rate $\Omega$-LPF-ADC block, and each of the remaining $M_2$ signals is sampled via a rate $\Delta$-LPF-ADC block.  Let $\mA_1$, and $\mA_2$ be the sub-matrices composed of the first $M_1$, and remaining $M_2$ rows of $\mA$, respectively, 
\begin{align}\label{eq:A-decomp}
	\mA = \begin{bmatrix} 
		\mA_1 \\
		 \mA_2
		\end{bmatrix}
\end{align}
where $M_1 +M_2 = M$.  Recall, we imagine $\mX_c(t)$ as a matrix containing the continuous time signal $\{x_m(t)\}_m$ as its rows. Then $\mA_1\mX_c(t) : = \{\tilde{x}_1(t),\ldots, \tilde{x}_{M_1}(t)\}$ are the top $M_1$ signals at the output of the AVMM. Each of these signals is multiplied by a binary waveform and the result is integrated over an interval of length $1/\Omega$, and the $n$th sample in the $m$th output signal is
\begin{align*}
& Y_1[m,n] = \int_{(n-1)/\Omega}^{n/\Omega}\tilde{x}_m(t)b(t)dt,  \text{where} \ m \in [M_1],  n \in [\Omega].
\end{align*}
As $b(t)$ is piecewise constant over intervals of length $1/W$, we can write the above integration as a summation 
\begin{equation}\label{eq:Entries-Y1}
Y_1[m,n] = \sum_{\ell \sim \setB_n} b[\ell] \int_{(\ell-1)/W}^{\ell/W}\tilde{x}_m(t)dt,
 \ m \in [M_1] \ n \in [\Omega], 
\end{equation}
where\footnote{We are implicitly assuming here that $\Delta \geq \Omega$, the modification of the proof for $\Delta \leq \Omega$ will be clear by the end. To reduce the clutter, we assume $\Omega$ as a factor of $W$; the argument can easily be modified when it is not the case.} $\setB_n := \{ (n-1)W/\Omega+1, (n-1)W/\Omega+2 , \ldots, nW/\Omega\},$ and $\ell \sim \setB_n$ is a shorthand for $\ell$ taking all the values in $\setB_n$. Define a matrix $\tilde{\mX}$ whose entries are 
\begin{align}\label{eq:tildeX-entries}
\tilde{X}[m,\ell] &= \int_{(\ell-1)/W}^{\ell/W}\tilde{x}_m(t)dt\notag\\
&= \sum_{\omega \in \setW}\widetilde{C}[m,\omega]\left[\tfrac{e^{\iota 2\pi \omega / W} - 1}{\iota 2\pi \omega}\right]e^{-\iota 2\pi\omega \ell/W}, 
\end{align}
where the second equality follows by using DFT expansion, and $\widetilde{\mC}$ are DFT coefficients of $\tilde{x}_m(t)$ defined in \eqref{eq:tildeC}. Define an $W \times W$ diagonal matrix $\mT$ with entries $T[\omega,\omega] = \left[(e^{\iota 2\pi \omega / W} - 1)/\iota 2\pi \omega\right]$. Matrix $\mT$ is invertible as $T[\omega,\omega] \neq 0$ for every $\omega \in \setW$. In matrix form, \eqref{eq:tildeX-entries} becomes 
\begin{align}\label{eq:tildeX}
\tilde{\mX} = \tilde{\mC}\mT\mF^* = \mA\mC\mT\mF^*.
\end{align}
Define an $(\alpha,\beta)$th entry of an $\Omega \times W$ matrix $\mP_{\Omega,W}$ as follows 
\begin{align}\label{eq:P}
P_{\Omega,W}[\alpha,\beta] = 
\begin{cases}
1 & \text{for every} \ (\alpha,\beta) \in (n,\setB_n) \ \text{and}\  n \in [\Omega]\\
0 & \text{otherwise}. 
\end{cases}
\end{align}
In words, $\mP_{\Omega,W}\vx$ returns a length $\Omega$ vector by summing $W/\Omega$ adjacent entries of $\vx$. In an exactly similar manner, we can also define $\mP_{\Delta, W}$, and $\mP_{\Delta,W}\vx$ collapses $\vx$ into a length $\Delta$ vector by summing $W/\Delta$ adjacent entries. Evidently, every entry of $\mY_1$ in \eqref{eq:Entries-Y1} is the sum of the a few entries of a row of $\tilde{\mX}$ scaled by binary numbers $b[\ell]$'s. In light of \eqref{eq:tildeX}, equation \eqref{eq:Entries-Y1} in matrix form is $\mY_1 = \mA_1\mC\mT\mF^*\mD^*\mP_{\Omega,W}^*,$ where $\mD = \text{diag}(b[1], b[2], \ldots, b[W])$ is a diagonal matrix. 

Samples collected in the bottom $M_2$ branches can be expressed in matrix form using the same approach; the only difference is that in place of a rate $\Omega$-LPF-ADC block, we now have a rate $\Delta$-LPF-ADC block. Samples in the bottom $M_2$ branches are collected in a $M_2 \times \Delta$ matrix $\mY_2$ given by $\mY_2 = \mA_2 \mC\mT\mF^*\mD^*\mP^*_{\Delta,W}.$ To ease the notation, we define 
\begin{align}\label{eq:HandQs}
\mQ_1 &= \mP_{\Omega,W}\mD\mF, ~\mQ_2 = \mP_{\Delta,W}\mD\mF, ~\text{and}~ \mH = \mC\mT. 
\end{align}
Observe that $\mH$ inherits rank-$R$, and $S$-sparse-rows structure from $\mC$. Our objective of recovering the unknown $\mH$  from a few linear measurements $\mY_1 = \mA_1\mH\mQ_1^*, \ \mY_2 = \mA_2 \mH\mQ_2^*$ leads to an under-determined system of equations. Among multiple candidates of solution, in this case, we choose the one with S\&L structure. To enforce this, a natural way is to solve an $\ell_1$-plus-nuclear-norm penalized semidefinite program. In the  general case of noisy measurements 
\begin{align}\label{eq:Y1-Y2}
\mY_1 = \mA_1\mH\mQ_1^* + \mE_1, \ \mY_2 = \mA_2 \mH\mQ_2^*+\mE_2,
\end{align}
where the additive matrices $\mE_1$, and $\mE_2$ account for the bounded ($\|\mE_1\|_{\F} \leq \delta_1$, and $\|\mE_2\|_{\F} \leq \delta_2$) measurement noise, the semidefinite program becomes
\begin{align}\label{eq:semidefinite-program}
&\underset{\mH}{\text{minimize}} ~\|\mH\|_*+\lambda \|\mH\|_{1,2}\\
&\text{subject to} ~  \|\mY_1 - \mA_1\mH\mQ_1^*\|_{\F} \leq \delta_1\notag\\
& \qquad \qquad \ \|\mY_2  = \mA_2 \mH\mQ_2^*\|_{\F}\leq \delta_2,\notag
\end{align}
where the $\ell_{1,2}$, and nuclear-norm penalties favor the column sparse, and low-rank solutions, respectively, and $\lambda \geq 0$ is a free parameter. However, the optimization program in \eqref{eq:semidefinite-program}, or any other objective involving a combination of both these norms does not yield an effective penalty for S\&L matrices as it provably  fails \cite{oymak2015simultaneously} whenever
\[
\text{Total \# of measurements} \lesssim c\min(MS,RW).
\]
In other words, one needs at least a sampling rate $\setO(MS,RW)$ --- which is much smaller than the Nyquist rate $MW$ but still potentially much larger than the optimal rate $RS$, derived from the underlying number $RS$ of unknowns in $\mH$ --- to have any possibility of signal recovery. 

Moreover, the semidefinite program is computationally expensive, and it quickly becomes impractical to solve this for medium scale values of $M$, and $W$.  The main reason is the unknowns in \eqref{eq:semidefinite-program} scale with $MW$, and not with the actual number $RS$ of unknowns.   We, therefore, devise a different approach to recover $\mH$ by first cheaply finding the $R$ basis vectors for each of the row (left), and column (right) space, and following it up with a simple least squares program to recover the smaller $R\times R$ intermediate matrix. 

\section{Column and Row Space Measurements}\label{sec:Column-Row-Measurements}
Our strategy to solve for $\mH$ relies on the observation that if the bases of the column and row space of $\mH$ are known then its recovery reduces to solving a simple least squares program \cite{ahmed2017compressive,halko2011finding}. In this section, we extract column and row space bases of $\mH$ from the observed samples $\mY_1$, and $\mY_2$. 

Verify using the definition in \eqref{eq:P} that\footnote{To avoid deviating from the main point, and to reduce the clutter, we restrict ourselves to the case when $\Omega$ is a factor of $\Delta$. Again modification to the general case is easy.} $\mP_{\Omega,\Delta}\mP_{\Delta, W} = \mP_{\Omega, W}$. The column measurements of $\mH$ can be extracted from $\mY_1$, and $\mY_2$ in \eqref{eq:Y1-Y2} as follows
\begin{align}\label{eq:Yc}
\mY_c &= \mA^*\begin{bmatrix}
\mY_1\\
\mY_2\mP_{\Omega,\Delta}^* 
\end{bmatrix}
 = \mA^*\mA\mH\mQ_1^* + \mA^* \begin{bmatrix} \mE_1 \\
\mE_2 \mP_{\Omega,\Delta}^*
\end{bmatrix}\notag\\
&= \mH\mQ_1^* + \mE_c,
\end{align} 
where last equality follows from the fact that $\mA^*\mA = \mI$, and $\mE_c := \mA^* \begin{bmatrix} \mE_1 \\
\mE_2 \mP_{\Omega,\Delta}^*
\end{bmatrix}$. Using the fact that $\|\mP_{\Delta,\Omega}\| = \sqrt{\Delta/\Omega}$, it is easy to see that 
\begin{align}\label{eq:Ec-bound}
\|\mE_c\|_{\F} \leq \|\mE_1\|_{\F} + \sqrt{\frac{\Delta}{\Omega}}\|\mE_2\|_{\F} \leq \delta_1 + \delta_2\sqrt{\frac{\Delta}{\Omega}}.
\end{align} 
The name column-space measurements for $\mY_c$ comes from the fact that columns of the matrix $\mH\mQ_1^*$  are random linear combinations of the columns of $\mH$, and hence serve as samples of column space of $\mH$. Using a similar reasoning, $\mA_2\mH$ are the row-space measurements of $\mH$. Unlike directly observing column measurements $\mH\mQ_1^*$ in $\mY_c$, we do not observe the row-space measurements $\mA_2\mH$ directly but only a random projection $\mY_2 = \mA_2\mH\mQ_2^*$ of the row-space measurements through an under-determined random projection operator $\mQ_2$. 
\section{Signal Reconstruction Algorithm}\label{sec:Algo}
Recall that $\mH$ has at most $S$-sparse rows with common support; please refer to the signal model in Section \ref{sec:Signal-Model}. This means $\mA_2\mH$ also has at most $S$-sparse rows, and to recover an estimate of  row-space measurements $\mA_2\mH$ from its under-determined set of linear observations $\mY_2$ in \eqref{eq:Y1-Y2}, we solve an $\ell_1$ minimization program: 
 \begin{align}\label{eq:l1-program}
\mY_r := &\underset{\mZ \in \C^{M_2 \times W}}{\text{argmin}}~ \|\mZ\|_1 \ \text{subject to} \ \|\mY_2 - \mZ\mQ_2^* \|_{\F} \leq \delta_2,
\end{align}
where the estimate $\mY_r$ is intended to be used as the row space measurements. 
 
 We now take the top $R$ left singular vectors $\mL_R$ of $\mY_c$ in \eqref{eq:Yc} as the basis of the column space of $\mH$. The estimate $\hat{\mH}$ of $\mH$ is then formed as
\begin{align}\label{eq:hatH}
\hat{\mH} = \mL_R \mS
\end{align}
for an unknown $R \times W$ matrix $\mS$, which is obtained by solving the following least-squares  program using the row-space samples $\mY_r$ in \eqref{eq:l1-program} as follows
\begin{align}\label{eq:least-squares-program}
\mS  := \underset{\mZ \in \C^{R \times W}}{\text{argmin}} ~ \|\mY_r - \mA_2\mL_R \mZ\|^2_{\F}.
\end{align}
A closed-form solution of this program is simply 
\[
\mS = (\mA_2\mL_R)^\dagger \mY_r, 
\]
where $\dagger$ denotes the pseudo inverse. 

Recall that $\mH= \mC\mT$. Given the estimate of $\mH$ in \eqref{eq:hatH}, an estimate of the Nyquist rate samples $\mX$ in \eqref{eq:X-def} is obtained using $\hat{\mX} = \hat{\mH}\mT^{-1}\mF^*$. The signal ensemble $\mX_c(t)$ can then be determined using the conventional linear sinc interpolation. 
\section{Coherence}

Our results show that a sufficient compressive sampling rate to recover the signal ensemble also depends on the dispersion of signals across time. Since the compressive sampling rate is potentially far fewer than the Nyquist rate, the ADCs can end up sensing mostly zeros for a signal that is localized across time. Ideally, we want the signals to be well-dispersed across time to recover them from as few compressive samples as possible. This intuition is also supported by Theorem \ref{thm:main}, which shows that the sufficient sampling rate scales with a coherence parameter $\mu_0^2$, defined below. 

Let  $\mH= \mU\mSigma\mV^*$  be the SVD of $\mH$, and recall that the rows of $\mH$ are the modified (low-pass filtered) frequency spectrum of the signals in the ensemble, respectively. The best rank-$R$ approximation of $\mH$ is 
\begin{align}\label{eq:HR}
\mH_R = \mU_R\mSigma_R\mV_R^*,
\end{align}
where $\mU_R$ are the top $R$ columns of $\mU$, and $\mV_R$ is defined similarly. $\mSigma_R$ is the $R \times R$ matrix of top $R$ singular values. Our theoretical results show that the sampling rate scales with a coherence parameter defined as
\begin{align}
\mu_0^2 : = \frac{W}{R}\|\mF\mV_R\|^2_{2 \rightarrow \infty},
\end{align}
where $\|\mF\mV_R\|_{2 \rightarrow \infty}$ norm returns the maximum of the $\ell_2$-norms of the rows of $\mF\mV_R$, and $\mF$ is defined in \eqref{eq:DFT}. The coherence can be best understood by relating $\mu_0^2$ to $\|\mH_R\mF^*\|_{2 \rightarrow \infty}$ --- the \textit{collective} peak value of the signal ensemble across time. For a fixed energy ensemble, the smaller value of this quantity means a more dispersed across time and vice versa. It is easy to check that $1 \leq \mu_0^2 \leq W/R$. To see this, let $\vf_\ell^*$ be the $\ell$th row of $\mF$, we can write $\|\mF\mV_R\|^2_{2 \rightarrow \infty} = \max_{\ell} \| \vf_\ell^*\mV_R\|_2^2$. This implies that 
\begin{align*}
W\mu_0^2 \geq \frac{W}{R} \sum_{\ell=1}^W \|\vf_\ell^*\mV_R\|_{2}^2 = \frac{W}{R} \|\mF\mV_R\|_F^2 = W.
\end{align*}
This gives $\mu_0^2 \geq 1$. In addition, $\|\vf_\ell^*\mV_R\|_2^2 \leq \|\mV_R\|^2\|\vf_\ell\|_2^2 \leq 1$, where $\|\cdot\|$ is the operator norm. This gives $\mu_0^2 \leq W/R$. 
Smallest, and largest values correspond to perfectly flat, and very spiky signals across time, respectively.
 
 Additional preprocessing using random filters to force signal diffusion across time can be added in the sampling architecture \cite{ahmed2015compressive2}. This leads to sampling rates that are independent of the coherence parameter $\mu_0^2$. 

\section{Sampling Theorem}
Let $\mH_R$ be the best rank-$R$ approximation of $\mH$ as in \eqref{eq:HR}, and $(\mA_2\mH)_S$ be the best $S$-row-sparse (all the rows are at most $S$-sparse) approximation of $\mA_2\mH$ in the conventional Frobenius norm. We now state a sampling theorem showing that the signal ensemble $\mX_c(t)$ can be recovered exactly in the noiseless case and stably in the noisy case via the proposed reconstruction algorithm in Section \ref{sec:Algo}. 

\begin{thm}\label{thm:main}
	Given the samples $\mY_1$, and $\mY_2$ of the unknown matrix $\mH$, as defined in \eqref{eq:HandQs} that are contaminated with $\delta_1,\delta_2$ bounded noise, as constructed in  \eqref{eq:Y1-Y2}.  Let $\mA$ be a random orthogonal matrix as in \eqref{eq:ATA} and \eqref{eq:A-decomp}. The estimate $\hat{\mH}$ in \eqref{eq:hatH}, obtained by solving $\ell_1$-program in \eqref{eq:l1-program} followed by a least-squares in \eqref{eq:least-squares-program} obeys
	\begin{align}\label{eq:exact-stable-recovery}
	&\|\hat{\mH}-\mH\|_{\F} \leq  c\sqrt{\frac{M}{M_2}}\Bigg[\sqrt{\frac{W}{\Omega}} \|\mH-\mH_R\|_{\F} + \notag \\
	&\quad \delta_1 + \sqrt{\frac{\Delta}{\Omega}}\delta_2 + \frac{1}{\sqrt{M_2S}}\|\mA_2\mH- (\mA_2\mH)_S\|_{1} \Bigg]
	\end{align}
	with probability at least $1-\setO(W^{-\beta})$ whenever $\Delta \geq C_\beta S\log^6 W$, $M_2 \geq C(R+\beta \log W)$, $M \geq C(M_2+\beta\log W)$, $\Omega \geq C_\beta \mu_0^2 R \log^2 W$, and $\Delta \geq \Omega$. 
\end{thm}
\textit{Proof:} We defer the proof of Theorem 1 to Section \ref{sec:Proof}.
\subsection{Discussion on Theorem \ref{thm:main}}
In the sampling architecture shown in Figure \ref{fig:Sampling-Architecture}, the ADCs in the top $M_1$ channels ($M = M_1+M_2$) operate at a rate $\Omega$, and the remaining $M_2$ channels operate at a rate $\Delta$.  Theorem \ref{thm:main} implies that it suffices to set the cumulative-sampling rate (CSR) for signal reconstruction\footnote{The notation $A \gtrsim B$ means that $ A \geq c B$ for an absolute constant $c$.} at 
\begin{align*}
\text{CSR}=\Omega M_1 + \Delta M_2 \gtrsim M R \log^2 W +RS\log^6 W
\end{align*}
assuming that $M_1 \gtrsim \beta \log W$, $R \gtrsim \beta \log W$, $M_2 = \setO(R)$, and signals are well dispersed across time ($\mu_0^2 \approx 1$). In practical applications, the effective signal bandwidth $S$ is much more than the number $M$. In this case, the net sampling rate roughly simplifies to more readable form: 
\begin{align*}
\Omega M_1 + \Delta M_2 \gtrsim RS\log^6 W. 
\end{align*}
Compare this rate to the Nyquist rate of $MW$ samples per second. Evidently, this results in significant reduction in sampling rates when signals are correlated $R \ll M$, and spectrally sparse $S \ll W$. 

Finally, exact recovery result follows from Theorem \ref{thm:main} in the noise less case $\delta_1 = 0$, and $\delta_2 = 0$, the ensemble is exactly $S$-row sparse giving $\mA_2\mH = (\mA_2\mH )_S$, and is also exactly rank-$R$ giving $\mH = \mH_R$. Plugging these in \eqref{eq:exact-stable-recovery} shows that $\hat{\mH} = \mH$ in this case. 

\subsection{Choosing $M_1$ and $M_2$}\label{sec:M1-M2-choice}
 Our choice of feasible number $M_1$ of channels in which ADCs operate at a rate $\Omega$, and feasible number $M_2$ of bottom channels in which ADCs operate at a rate $\Delta$ must conform to $M = M_1+M_2$, and $M_2 \gtrsim R + \log W$,  as required by Theorem \ref{thm:main}. 
 What is a good choice of the decomposition $M = M_1+M_2$ to minimize the cumulative sampling rate $M_1\Omega + M_2\Delta$? Since $\Delta \geq \Omega$, we must choose $M_2$ to be a smallest feasible number.

The choice of $\Delta$ and $\Omega$ is in turn dictated by $S$ and $R$ as stated in Theorem \ref{thm:main}.

\section{Related Work} 

Exploiting inherent signal structures such as spectral sparsity and correlation to achieve gains in sampling rate has been actively studied \cite{tropp2010beyond,slavinsky11co,ahmed2015compressive2} after the advent of compressive sensing \cite{candes2008introduction}. 
New sampling theorems proving the sub-Nyquist acquisition of spectrally-sparse signals have been rigorously established using the tools and ideas developed in the vast literature of sparse signal processing \cite{starck2010sparse}. The central idea is to diffuse the analog signals with preprocessing before sampling at a lower rate. The analog preprocessing is handled in real time using implementable sampling architectures. In \cite{tropp2010beyond}, authors propose a sampling architecture that modulates a signal of bandwidth $W/2$ but with only $S$ active frequency components, where $S \ll W$. %Modulation is a pointwise multiplication of the signal with a random binary waveform in time. 
This smears the information content across the entire bandwidth and enables a following ADC to operate at a sub-Nyquist rate of only $S\log^{\alpha} W$, where $\alpha$ is a known small constant. A digital post-processing using an $\ell_1$-minimization program provably reconstructs the original signal from the acquired compressive samples. Multiple spectrally sparse signals can also be mixed and acquired using a single low-rate ADC. From this information individual signals can be untangled and recovered using sparse digital post-processing. Similar ideas are extended, and actual sampling architectures are implemented on chip for multiband signals; see, for example, \cite{mishali11xasub,mishali2009blind,wang2018phased}.

Correlation structure in an ensemble of signals has also been effectively used to lower the sufficient sampling rate potentially way below the Nyquist rate. In a nutshell, the proposed sampling schemes in \cite{ahmed2011compressive,ahmed2012compressive,ahmed2013compressive,ahmed2015compressive,ahmed2015compressive2,ahmed2017compressive} can acquire the signal ensemble $\mX_c(t)$ above at a rate of $RW\log^{\alpha} W$, which is potentially much smaller than the Nyquist rate $MW$  when $R \ll M$. The signal reconstruction problem in this case can be framed as a recovery of an $M \times W$ matrix of rank $R$ from an under-determined set of linear measurements, which can  be effectively solved using a nuclear-norm penalized semidefinite program. Nuclear-norm penalty enforce low-rank structure on the unknown matrix, which effectively exploits the correlation in the signal ensemble. Implementable sampling architectures for individual, and multiplexed signals are presented in detail in \cite{ahmed2011compressive,ahmed2013compressive,ahmed2015compressive2}, and \cite{ahmed2012compressive,ahmed2015compressive} along with a rigorous development of the related sampling theorems. 

 The prior art on signal reconstruction from sub-Nyquist rate samples mostly considers either sparse or correlated signal structure on the signal ensemble. We consider signal ensembles that are simultaneously sparse and correlated. We frame the signal reconstruction as a sparse-and-low-rank matrix recovery problem from an under-determined set of linear measurements. A naive extension of simply using the combination $\ell_1$ and nuclear norm penalties is not effective in this case as shown in \cite{oymak2015simultaneously}. A more detailed comparison with \cite{oymak2015simultaneously} was already discussed in Section \ref{sec:DiscreteForm}. We develop a novel two-step recovery algorithm that solves an $\ell_1$ minimization program followed by a simple least squares program to recover a stable estimate of the ground truth from an optimal (within log factors) sampling rate of $RS\log^{\alpha}W$. Using earlier works \cite{tropp2010beyond, ahmed2015compressive,ahmed2015compressive2} that can only take advantage of either sparse or correlation structure in the signal ensemble, one requires a sub-optimal sample rate $\min(RW,MS)\log^{\alpha}W$ to reconstruct the S\&C ensemble $\mX_c(t)$, whereas in comparison we only require a potentially much smaller rate $RS\log^{\alpha}W$ as $S \ll W$, and $R\ll M$.  Moreover, we reconstruct the signal with a computationally much less expensive algorithm compared to the semidefinite program above.

\section{Applications}\label{sec:Applications}

One application area in which sparse and correlated signals play a central role is array processing. High-density arrays with hundreds to thousands of array elements are increasingly being employed in phased-array automotive radars \cite{ku201316, alhalabi201477, sridharan1998us,maskell2008sapphire,miller2007new}. %for space surveillance and tracking to efficiently monitor increasing amount of satellites, and space debris, on-chip integrated phased arrays using, e.g., piezoelectric devices for biosensing \cite{bowen1993injection}, and in micro electrode arrays (MEA) to study the generation and propagation of neuronal action potentials \cite{frey07ce,imfeld08la,haas08pr,gray04di}. Another important domain is robotics, where high density tactile sensor arrays using pressure and temperature sensors integrated on-chip \cite{tenzer2014inexpensive} with analog-to-digital converters, and amplifiers, etc. These sensor arrays are used in a wide range of human, and environment interface applications, e,g., a robotic hand for grasping, and manipulation. 
Signals recorded by such massive numbers of sensors/array elements often have a lot of spatial and temporal redundancies that are well modeled by an S\&C ensemble, which can then be exploited to obtain potentially significant reductions in the required sampling rate using the proposed sampling scheme. This leads to a reduction in the huge volume of data generated in the automotive radar application, less power dissipation, and comparatively cheaper, and more precise analog-to-digital converters. 

\begin{figure*}
	\centering
	\begin{tabular}{cc}
		\includegraphics[height=1.9in]{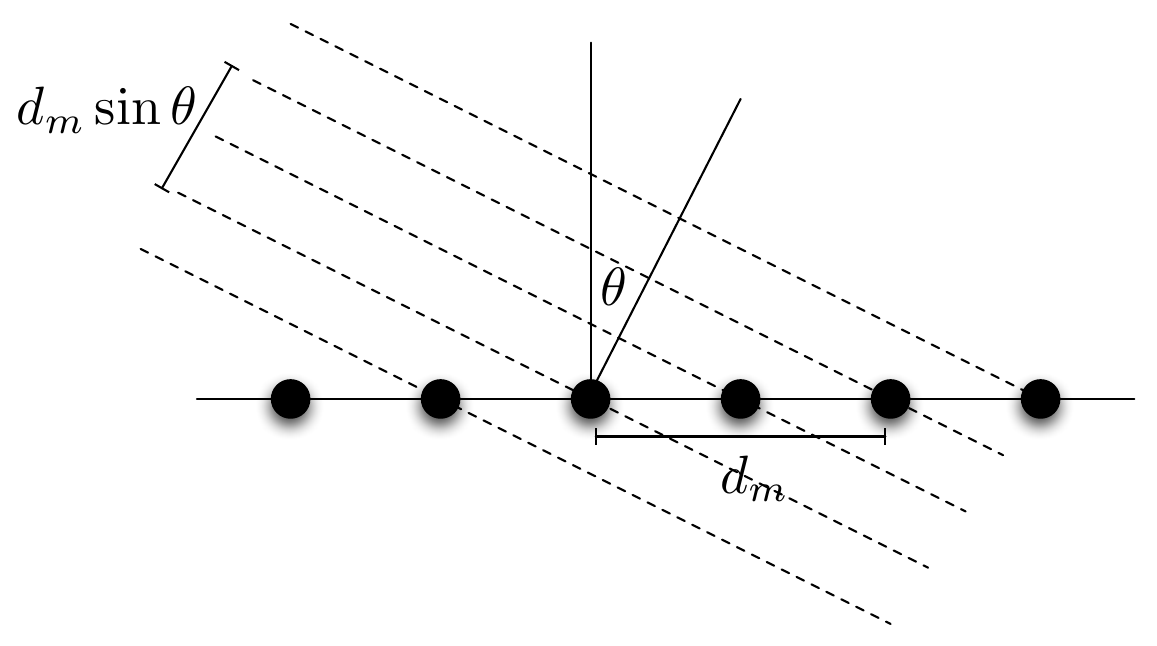} \hspace*{0.1in} &		
		\raisebox{0.2in}{\rotatebox{90}{\small $\log_{10}(k$th largest eigenvalue$)$}}
		\includegraphics[height=1.9in]{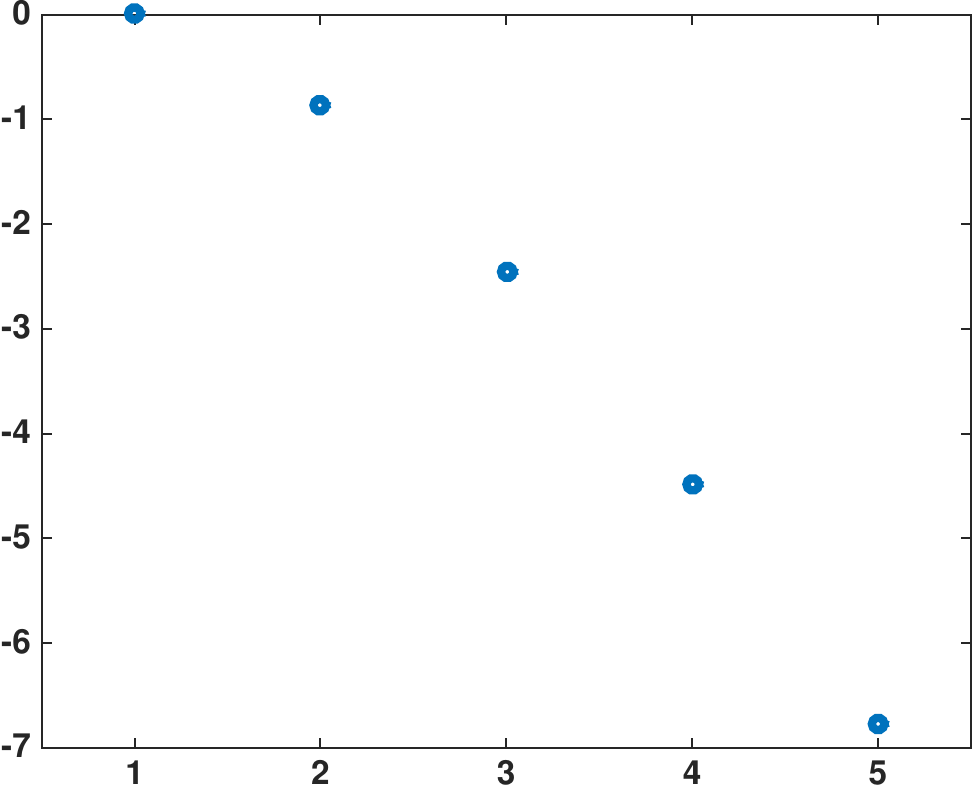} \\[-2mm]
		& $k\rightarrow$ \\
		(a) & (b) 
	\end{tabular}
	\caption{\small\sl (a) A plane wave impinges on a linear array in free space.  When the wave is a pure tone in time, then the responses at each element will simply be phase shifts of one another.  (b) Eigenvalues for $\mR_{aa}$, on a $\log_{10}$ scale and normalized so that the largest eigenvalue is $1$, defined in \eqref{eq:Raa} for an electromagnetic signal with a bandwidth of $100$ MHz and a carrier frequency of $5$ GHz; the array elements are spaced half a carrier-wavelength apart.  Even when the signal has an appreciable bandwidth, the signals at each of the array elements are heavily correlated --- the effective dimension, in this case, is $R=3$ or $4$.}
	\label{fig:arrayprocessing}
\end{figure*}

The central theme is that multiple signals are emitted from different locations. Each signal sparsely occupies a bandwidth $W$, and is modulated up to a carrier frequency $\omega_c$. A single tone signal $\mathrm{e}^{\iota 2 \pi \omega t}$  arrives at multiple array elements record signals with different time shifts, determined by the spacing between array elements  as illustrated in Figure \ref{fig:arrayprocessing}.  As an illustration, the signal arriving at the $m$th array element of an $M$-element array in the simple case of a single emitter is
\begin{align}\label{eq:single-emitter}
x_m(t) =  \int_{\omega_c-W/2}^{\omega_c+W/2}\mathrm{e}^{-\iota 2 \pi \omega d_m \sin \theta/c}\alpha_{m,\omega}\mathrm{e}^{\iota 2 \pi \omega t}d\omega, 
\end{align}
where $\alpha_{m,\omega}\mathrm{e}^{-\iota 2 \pi \omega d_m \sin \theta/c}: = a_m(\theta,\omega)$ is referred to as steering gain at the $m$th array element, where $\mathrm{e}^{-\iota 2 \pi \omega d_m \sin \theta/c}$ is the phase shift caused in the $\omega$ frequency tone due to the arrival delay $\tau_m = d_m \sin \theta /c$, and $\alpha_{m,\omega}$ is the gain or strength of $\omega$ tone at $m$th array element.   The integral simply aggregate the contributions of frequency components present in the entire bandwidth $W$. The signal ensemble $\mX_c(t)$ is the stack of $x_m(t), 1\leq m \leq M$ as its rows\footnote{The elements of the rows are the samples $x_m(t_\ell)$ in a given window of time.}. This gives
\begin{align*}
\mX_c(t) = \int_{\omega_c -W/2}^{\omega_c + W/2}\va(\theta, \omega)\mathrm{e}^{\iota 2 \pi \omega t} d\omega, 
\end{align*}
where the length $M$ column $\va(\theta,\omega)$ is the \textit{steering} vector. Evidently,  $\va(\theta, \omega)\mathrm{e}^{\iota 2 \pi \omega t}$ is a rank-one ensemble, where we think of the signal $\mathrm{e}^{\iota 2 \pi \omega t}$ as a row vector obtained after eventual sampling across time $t$. The ensemble $\mX_c(t)$ is obtained by integrating the rank-one ensembles over the narrow-band $W$. The conceptual approach is exactly the same even in the case of multiple emitters as the steering vector $\va(\vtheta,\omega)$ is now a function of multiple incident angles, stacked in a vector $\vtheta$, due to wavefronts from different emitters. However, even in this case the quantity is $\va(\vtheta, \omega)\mathrm{e}^{\iota 2 \pi \omega t}$ is still a rank-one ensemble.

The only question that remains to be determined is how the integration over the bandwidth $W$ increases the rank. The answer to this question depends on the density of the array elements compared to the bandwidth $W$. We will show that for narrow-band signals, and high-density arrays, the rank of $\mX_c(t)$ remains low. Having an array with a large number of appropriately spaced elements can be very advantageous even when there are only a relatively small number of emitters present.  Observing multiple delayed versions of a signal allows us to perform spatial processing,  we can beamform to enhance or null out emitters at certain angles, and separate signals coming from different emitters.  The resolution to which we can perform this spatial processing depends on the number of elements in the array (and their spacing).  For high-density antenna arrays or narrow-band signals, the spatial sampling rate $1/\tau_m$ is much larger than the bandwidth $W$. This gives rise to very correlated steering vectors $\va(\theta, \omega)$. In the standard scenario, where the array elements are uniformly spaced $c/(2\omega_c)$ along a line, we can make this statement more precise using classical results on spectral concentration \cite{slepian76ba,slepian78pr}.  In this case, the steering vectors $\va(\theta,\omega)$ for $\omega\in [\omega_c\pm W/2]$ are equivalent to integer spaced samples of a signal whose (continuous-time) Fourier transform is bandlimited to frequencies in $(1\pm W/(2\omega_c))(\sin\theta)/2$, for a bandwidth less than $W/(2\omega_c)$.  Thus the dimension of the subspace spanned by $\{\va(\theta,\omega),~\omega\in[\omega_c\pm W/2]\}$ is, to within a very good approximation, $\approx MW\tau_m+ 1 = MW/\omega_c + 1$. 

Figure~\ref{fig:arrayprocessing}(b) illustrates a particular example.  The plot shows the (normalized) eigenvalues of the matrix
\begin{equation}
\label{eq:Raa}
\mR_{aa} = \int_{\omega_c-W/2}^{\omega_c+W/2} \va(\theta,\omega)\va(\theta,\omega)^*\, d\omega,
\end{equation}
for the fixed values of $\omega_c = 5$ GHz, $W = 100$ MHz, $c$ equals the speed of light, $M=101$, and $\theta=\pi/4$.  We have  $MW/\omega_c + 1 = 3.02$, and only $3$ of the eigenvalues are within a factor of $10^4$ of the largest one.

The correlated signal structure established on the input ensemble is well-known, and many spatial processing tasks, for instance, standard subspace methods \cite{schmidt86mu,roy1989esprit} for estimating the direction of arrival involve forming the spatial correlation matrix by averaging in time,
\[
\mR_{xx} = \frac{1}{L}\sum_{\ell=1}^L \mX(t_\ell)\mX(t_\ell)^*.
\] 
As the column space of $\mR_{xx}$ should be $\va(\theta,\omega)$, we can correlate the steering vector for every direction to see which one comes closest to matching the principal eigenvector of $\mR_{xx}$.  

The main results of this paper do not give any guarantees about how well these spatial processing tasks can be performed.  Rather, they say that the same correlation structure that makes these tasks possible can be used to lower the net sampling rate over time.  The entire signal ensemble can be reconstructed from this reduced set of samples, and spatial processing can follow.

On the other hand, the spectral sparsity of the ensemble $\mX_c(t)$ is controlled by the active frequencies in the bandwidth $W$, or, more precisely, the joint frequency band occupation of the emitters. It's easy to imagine several scenarios in practice in automotive radars \cite{feinberg2019hardware} where the frequency spectrum of the emitters is only sparsely occupied with a priori unknown support. Sparse frequency occupation can also be introduced, for example, when emitters transmit in disjoint frequency bands, and only a subset of the emitters are active at a given time.

It is fair, then, to say that the rank of the signal ensemble is a small constant time the number of narrow band emitters, and each array element can be easily imagined to be recording a very sparsely occupied signal spectrum.

\section{Numerical Experiments}\label{sec:Numerics}
In this section, we numerically simulate the reconstruction of S\&L matrix $\mH$ from the given measurements $\mY_1$, and $\mY_2$ in \eqref{eq:Y1-Y2} using our proposed algorithm in Section \ref{sec:Algo}. We form a synthetic S\&L matrix $\mH$ by multiplying a tall $M \times R$ dense, and a fat $R \times W$ sparse random matrix. The entries of these random matrices are independent Gaussian. 

\begin{figure*}[t]
	\centering
	\subfigure[\scriptsize]{
		\includegraphics[trim={0.3cm 0.3cm 0.8cm 0.0cm},clip,scale=0.53]{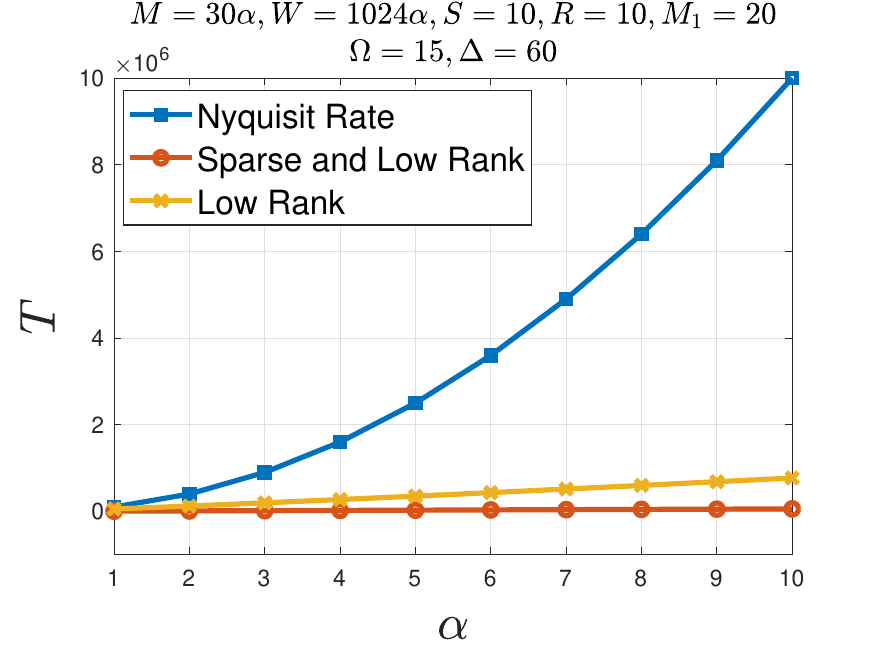}
    	\label{fig:T-vs-alpha}
    }
	\hspace{0.05cm}
	\subfigure[\scriptsize ]{
		\includegraphics[trim={0.1cm 0.3cm 0.8cm 0.0cm},clip,scale=0.53]{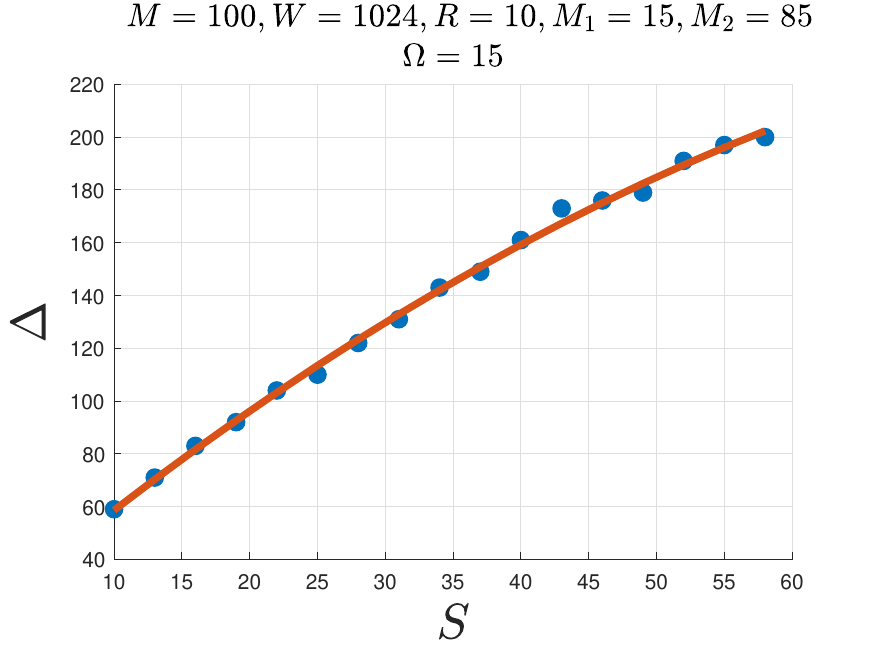}
	 	\label{fig:S-vs-Delta}
	} 
    \hspace{0.05cm}
	\subfigure[\scriptsize ]{
		\includegraphics[trim={0.1cm 0.3cm 0.8cm 0.0cm},clip,scale=0.53]{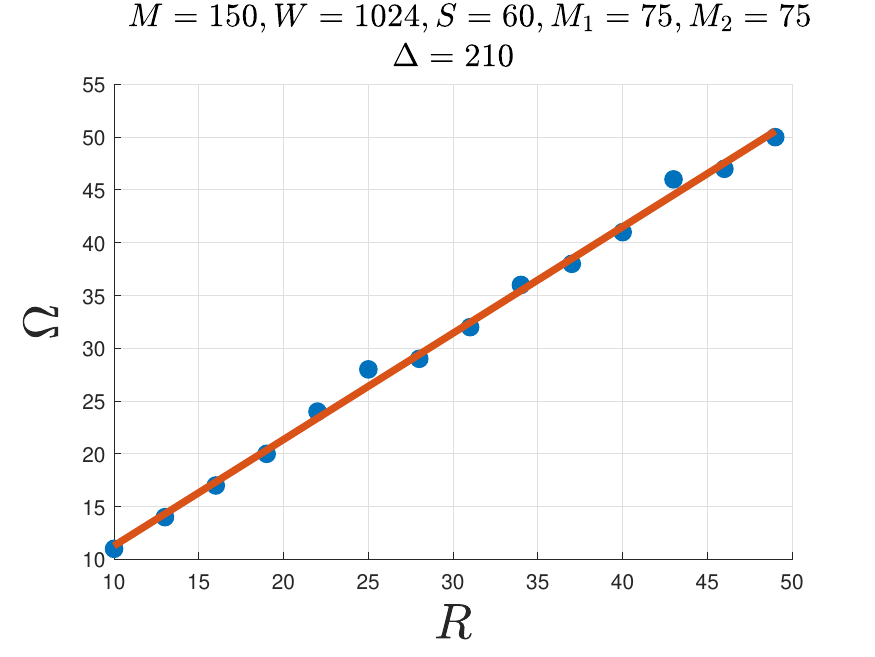}
		\label{fig:R-vs-Omega}
	}
	\caption{Illustrative plots between two parameters of interest from $M$, $W$, $S$, $R$, $M_1$, $\Omega$, and $\Delta$ while keeping the remaining fixed. (a)  $W$ and $M$ are set to increase linearly with $\alpha$. We plot $\alpha$ against minimum required cumulative sampling rate (CSR) for successful recovery of the signal ensemble using Nyquist criterion (blue), using reconstruction criterion of \cite{ahmed2015compressive2} (yellow), and using our proposed reconstruction criterion (orange). Nyquist rate quadratically increases with $\alpha$, the required sampling rate using the criterion of \cite{ahmed2015compressive2} only scales linearly with $\alpha$ as it only takes into account the correlated structure in the ensemble, and the sampling rate using our approach only scales very weakly (logarithmically) with $\alpha$ as it takes both sparse and correlated structure in the ensemble. (b) Spectral sparsity $S$ versus the minimum (required for the successful recovery of the ensemble) sampling rate  $\Delta$. As expected the sampling rate $\Delta$ (of an individual ADC in the bottom $M_2$ channels) scales linearly with $S$ and is much smaller than $W$. (c) Rank $R$ versus the minimum (required for the successful recovery of the ensemble)  sampling rate $\Omega$.  As expected the sampling rate $\Omega$ (of an individual ADC in the top $M_1$ channels) scales linearly with $R$ and is much smaller than $M$. 
		The discs in each case correspond to the minimum-sampling rate for signal reconstruction with an empirical success rate of a 99$\%$.
	  }
	\label{fig:SamplingRatePlots-1}
\end{figure*}

\begin{figure*}[h!]
	\centering 
	\subfigure[\scriptsize ]{
		\includegraphics[trim={0.1cm 0.3cm 0.8cm 0.0cm},clip,scale=0.52]{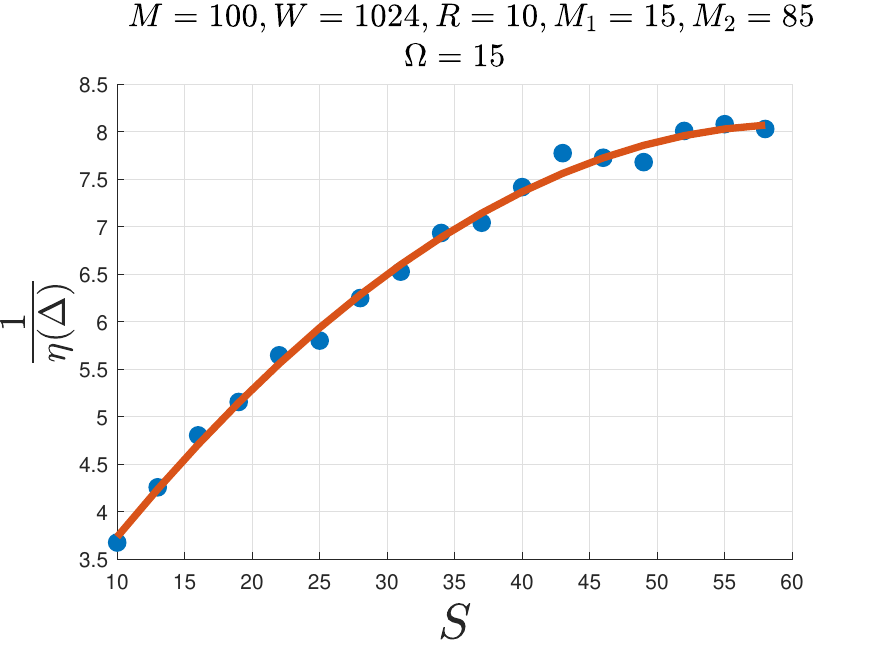}
	 \label{fig:S-vs-eta}
	 }
    \hspace{0.1in}
	\subfigure[\scriptsize ]{
		\includegraphics[trim={0.1cm 0.3cm 0.8cm 0.0cm},clip,scale=0.52]{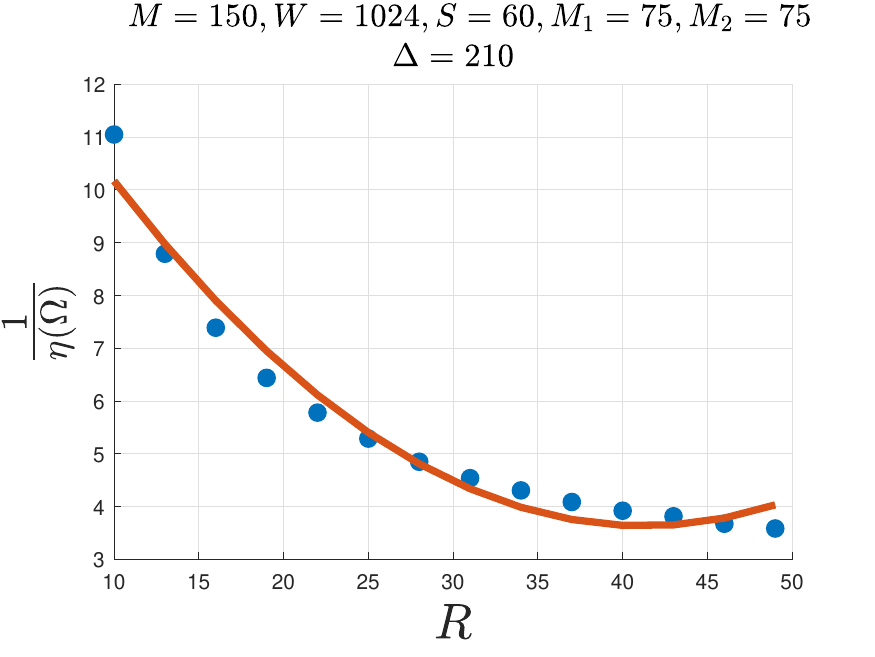}
		\label{fig:R-vs-eta}
	}
    \caption{(a) $S$ versus $1/\eta(\Delta)$. The sampling efficiency $\eta(\Delta)$ (a function of only one variable $\Delta$ keeping all other parameters fixed)  somewhat decreases with increasing $S$ but eventually settles down. (b) $R$ versus $1/\eta(\Omega)$.  The sampling efficiency $\eta(\Omega)$ (a function of only one variable $\Omega$ keeping all other parameters fixed)  somewhat increases with increasing $R$ and eventually settles down. The discs in left  plot correspond to the minimum value of $1/\eta(\Delta)$ as a function of the only parameter $\Delta$  that gives signal reconstruction with a 99$\%$ empirical success rate. Similar interpretation holds for discs in the right plot with respect to $1/\eta(\Omega)$ and $\Omega$. }
    \label{fig:SamplingRatePlots-2}
\end{figure*}
\begin{figure*}[h!] 
	\centering
	\subfigure[\scriptsize ]{\includegraphics[trim={0.12cm 0.14cm 0.3cm 0cm},clip,scale = 0.53]{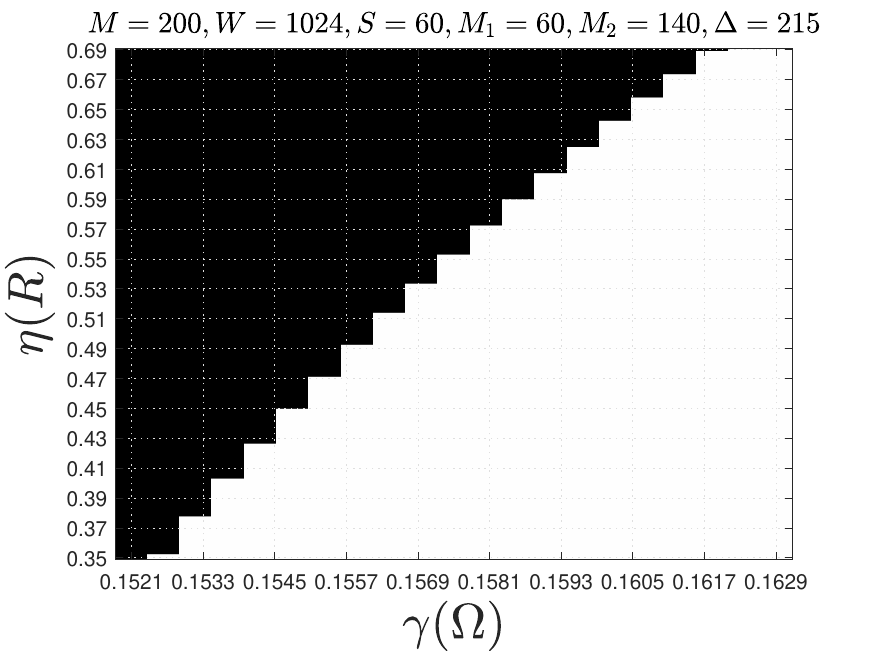}}\hspace{0.0in}
	\subfigure[\scriptsize ]{\includegraphics[trim={0.12cm 0.14cm 0.3cm 0.0cm},clip,scale = 0.53]{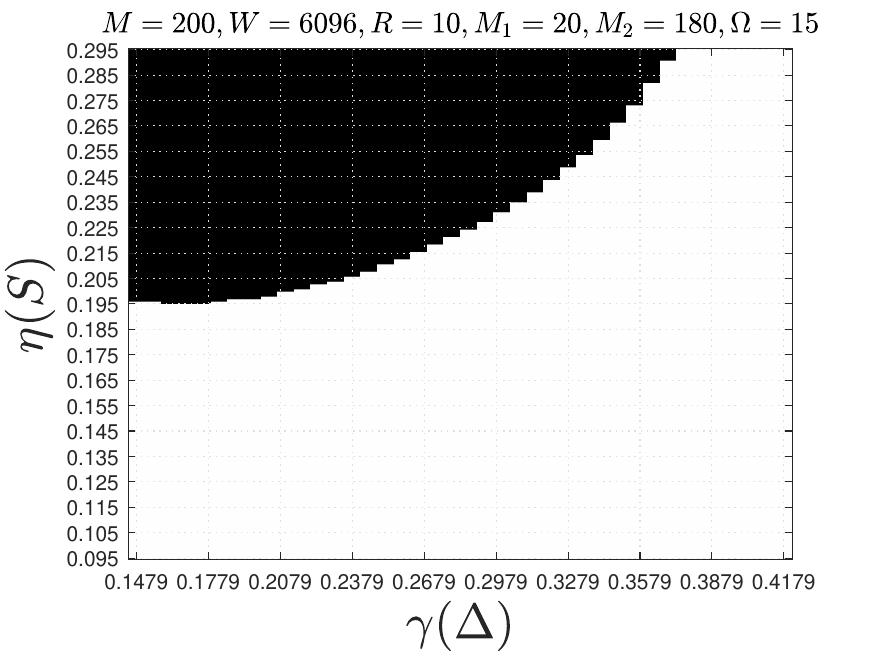}}
	
	\caption{Phase transitions between compression factor $\gamma$, and sampling efficiency $\eta$. Shade represents the probability of failure. (a) Phase Transitions between sampling efficiency $\eta(R)$ and $\gamma(\Delta)$ while keeping other parameters at fixed values, shown on the top; for example, $\Delta = 215$ is chosen in light of sparsity $S=60$ to avoid reconstruction failure. We then vary $R$, and $\Omega$ to obtain all the grid values of $\eta(R)$ and $\gamma(\Omega)$, and report the corresponding probability of failure at each grid point. Expectedly, increasing the sampling efficiency reduces the compression factor in the successful (white) region. (b) A similar phase transition between sampling efficiency $\eta(S)$ and $\gamma(\Delta)$.}
	\label{fig:phase_transition}
\end{figure*}

\begin{figure}[h!] 
	\centering
	\includegraphics[trim={0.25cm 0.16cm 1.2cm 0.3cm},clip,scale = 0.62]{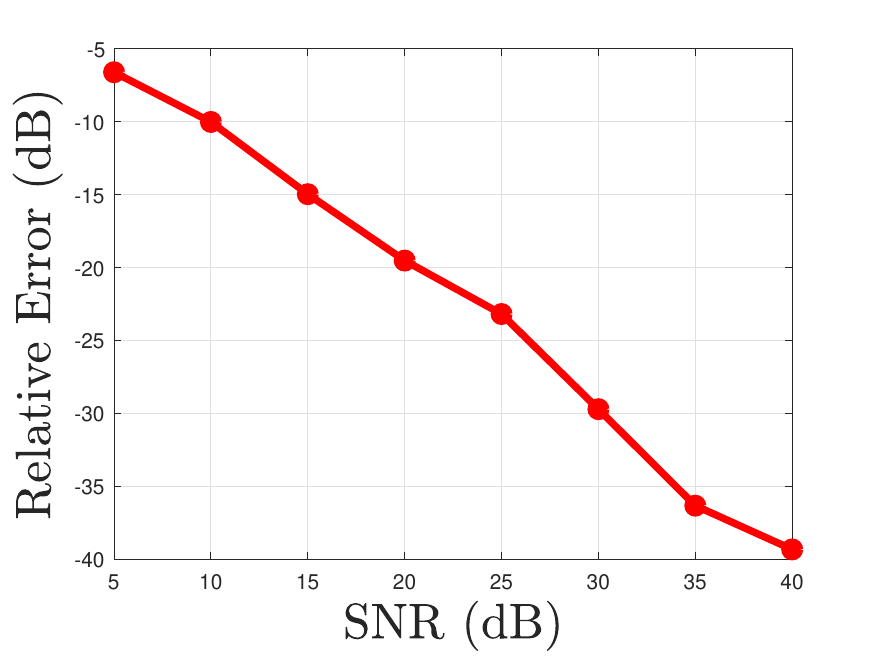}
	%\subfigure[\scriptsize ]{\includegraphics[width=.4\columnwidth]{untitled1.jpg}}
	\caption{ SNR (dB) vs. Relative error (dB).  We choose $M = 100$, $W/2 = 512Hz$, $S=10$, and $R =10$. Recovery using the proposed algorithm is stable in noise.}
	\label{fig:noise_results}
\end{figure}

We numerically evaluate the reconstruction algorithm by computing the relative error between the estimate $\hat{\mH}$ in \eqref{eq:hatH}, and the ground $\mH$ as follows
\begin{align}\label{eq:relative-error}
\text{relative error}:= \frac{\|\hat{\mH}-\mH\|_F}{\|\mH\|_F}.
\end{align}
In general, we declare a recovery $\hat{\mH}$ as successful whenever its relative error from the ground truth is less than $10^{-3}$. 

To facilitate the discussion, we also introduce the measures of sampling efficiency $\eta$, and compression factor $\gamma$ as follows
\begin{align*}
\eta := \frac{R(M+S-R)}{M_1\Omega + M_2\Delta} \  \text{and} \ \gamma := \frac{M_1\Omega + M_2\Delta}{MW}.
\end{align*}
The sampling efficiency $\eta$ is a ratio of the actual number of degrees of freedom in the unknown S\&L matrix $\mH$, and the cumulative number $T = M_1\Omega + M_2\Delta$ of linear measurements in $ \mY_1$ and $\mY_2$; see \eqref{eq:Y1-Y2}. In other words, sampling efficiency is the ratio between the minimum number of unknown parameters required to completely specify $\mX_c(t)$ in $t \in [0,1)$ and the cumulative sampling rate of the proposed scheme.  On the other hand, compression factor $\gamma$ is a ratio of the cumulative sampling rate and the Nyquist rate. Since $\eta$, and $\gamma$ are functions of multiple parameters; namely, $R$, $M$, $S$, $M_1$, $M_2$, $\Omega$, $\Delta$, and $W$. In our experiments, we will often vary $\eta$  by changing only one of these parameters such as $\Delta$, and fixing others, and use the notation $\eta(\Delta)$ to signify that $\eta$ is parametrized by $\Delta$ only while keeping others fixed to known values. Similarly, we will also use $\gamma(\Delta)$ or $\gamma(\Omega)$, etc. 

The first set of experiments in Figure \ref{fig:SamplingRatePlots-1} shows that successful reconstruction of S\&C signal ensemble can be numerically achieved using a rate much smaller than the Nyquist rate. For specific detail, please refer to the  image caption. 

The second set of experiments in Figure \ref{fig:SamplingRatePlots-2} show that the sampling efficiency $\eta(\Delta)$ vs. $S$ settles to $1/8$ after an initial small transition period. Similarly, the sampling efficiency $\eta(\Omega)$ vs. $R$ generally can be expected settles to as high as $1/4$ after an initial transition period. For more specific details on the experimental setup, please refer to the caption of the figure. 

The third set of experiments in Figure \ref{fig:phase_transition} show phase transitions between compression factor $\gamma(\Omega)$ and sampling efficiency $\eta(R)$; and between compression factor $\gamma(\Delta)$ and sampling efficiency $\eta(S)$. The shade shows the probability of failure; black is the failure probability of 1. We see that in both phase transitions that as the sampling efficiency increases, the compression factor decreases for successful reconstruction. For more specific details on the experimental setup, please refer to the caption of the figure. 

The fourth experiment  concerns reconstruction in the presence of noise. Figure \ref{fig:noise_results} plots SNR (dB) versus relative error (dB).  The relative error of the reconstructed ensemble degrades gracefully with reducing SNR.

\section{Conclusion}
In this paper, we propose a novel and implementable sampling architecture for the acquisition of a simultaneously sparse and correlated signal ensemble at a sub-Nyquist rate. The sampling architecture has applications in automotive radars. We prove a sampling theorem showing exact and stable reconstruction of the acquired signals even when the sampling rate is smaller than the Nyquist rate by orders of magnitude. The result of the sampling theorem has been validated via numerical simulations.

%\appendix
\section{Proof of Theorem \ref{thm:main}}\label{sec:Proof}
Recall that $\mL_R$ in \eqref{eq:hatH} are the top-$R$ left singular vectors of $\mY_c$, and let $\mV_R$ as in \eqref{eq:HR} be the top-$R$ right singular vectors of $\mH$. 
The proof relies on the upper, and lower bounds on the maximum, and minimum singular values, $\sigma_{\max}$ and $\sigma_{\min}$, respectively, of the matrices $\mQ_1\mV_R$, and $\mA_2\mL_R$. Lemma 1 in \cite{ahmed2017compressive} proves that for a fixed $\beta \geq 1$
\begin{align}\label{eq:singularvalue-bounds-Q1VR}
\sqrt{\frac{1}{2}}\leq \sigma_{\min} (\mQ_1\mV_R) \leq \sigma_{\max}(\mQ_1\mV_R) \leq \sqrt{\frac{3}{2}}
\end{align}
 with probability at least $1-\setO(W^{-\beta})$ whenever $\Omega \geq c\beta \mu_0^2 R \log^2 W$. As for $\mA_2\mL_R$, recall that $\mA$ in \eqref{eq:A-decomp} was assumed to be a random orthogonal matrix in \eqref{eq:ATA}. The orthogonality of $\mA$ has already been used in \eqref{eq:Yc}. Since $\mA_2$ is a fat random matrix with orthogonal rows, we can write 
 \[
 \mA_2= (\mG\mG^*)^{-1/2} \mG,
 \] 
 where $\mG$ is a standard Gaussian matrix; each entry is iid $\text{Normal}(0,1)$. The matrix $\mA_2\mL_R = (\mG\mG^*)^{-1/2} \mG\mL_R$, and $\mG\mL_R \sim \mG^\prime$, where $\mG^\prime$ is an $M_2 \times R$ standard Gaussian matrix. This simply means that 
 \[
 \sigma_{\max}(\mA_2\mL_R) \leq \sigma_{\max}(\mG^\prime) \sigma_{\min}^{-1}(\mG),
 \]
 and 
 \[
 \sigma_{\min}(\mA_2\mL_R) \geq \sigma_{\min}(\mG^\prime) \sigma_{\max}^{-1}(\mG).  
 \]
 Using standard result in random matrix theory; see, for example, Corollary 5.35 in \cite{vershynin10in},  the singular values of an $M_2 \times R$ Gaussian matrix $\mG^\prime$  obey 
 \[
 \sqrt{M_2/2} \approx \sigma_{\min}(\mG^\prime) \leq \sigma_{\max}(\mG^\prime) \leq \sqrt{2M_2}
 \]
  with probability at least $1-\setO(W^{-\beta})$ whenever $M_2 \geq c(R+\beta\log W)$ for sufficiently large constant $c$. Similarly, we have that 
  \[
  \sqrt{M/2} \approx \sigma_{\min}(\mG) \leq \sigma_{\max}(\mG) \leq \sqrt{2M}
  \]
   with probability at least $1-\setO(W^{-\beta})$ whenever $M \geq c(M_2 + \beta \log W)$. This directly implies that under the same conditions 
 \begin{align}\label{eq:singularvalue-bounds-A2UR}
 0.5 \sqrt{\frac{M_2}{M}}\leq \sigma_{\min}(\mA_2\mL_R)  \leq \sigma_{\max}(\mA_2\mL_R) \leq 2\sqrt{\frac{M_2}{M}}. 
 \end{align}
 Equation \eqref{eq:singularvalue-bounds-Q1VR}, and \eqref{eq:singularvalue-bounds-A2UR} directly imply that pseudo inverses $(\mQ_1\mV_R)^\dagger$, and $(\mA_2\mL_R)^{\dagger}$ are well defined, where 
 \begin{align}\label{eq:pseudo-inverse}
 (\mQ_1\mV_R)^{\dagger} = \big((\mQ_1\mV_R)^*(\mQ_1\mV_R)\big)^{-1}(\mQ_1\mV_R)^*, 
 \end{align}
 and similarly for $(\mA_2\mL_R)^{\dagger}$. 

 It is known that $\mQ_2$ obeys restricted isometry property (RIP) \cite{rudelson2008sparse,tropp2010beyond} over the set of sparse vectors. RIP then implies the exact and stable recovery of sparse rows of $\mA_2\mH$ using $\ell_1$-minimization program in \eqref{eq:l1-program}. Formally, Theorem 2 in \cite{tropp2010beyond} says that for a fixed $\beta \geq 1$ choose $\Delta \geq c_\beta S\log^6 W$ then with probability at least $1-\setO(W^{-\beta})$, the minimizer $\mY_r$ of the optimization program in \eqref{eq:l1-program} obeys 
\begin{align}\label{eq:Yr-stable-recovery}
\|\mY_r-\mA_2\mH\|_{\F} & \leq c^\prime \frac{1}{\sqrt{M_2S}}\|\mA_2\mH- (\mA_2\mH)_S\|_1 + c\delta_2,
\end{align} 
where $(\mA_2\mH)_S$ denotes the best approximation of the matrix $\mA_2\mH$ using $S$-sparse rows, and $c, c^\prime$ are fixed constants. Given $\mY_r$, the minimizer $\mS$ of the least squares program is simply $\mS = (\mA_2\mL_R)^{\dagger}\mY_r.$ 

We want to bound the distance of the estimate $\hat{\mH}$ in \eqref{eq:hatH} from the true $\mH$. Using triangle inequality, we have 
\begin{align}\label{eq:interim1}
\|\hat{\mH}-\mH\|_{\F} &\leq \|(\mI - \mL_R(\mA_2\mL_R)^\dagger\mA_2)\mH\|_{\F}\notag\\
& \qquad + \|\mL_R(\mA_2\mL_R)^{\dagger}(\mY_r - \mA_2\mH)\|_{\F}. 
\end{align}
For brevity, we denote $\mB = \mK \mA_2$ where $\mK =\mL_R(\mA_2\mL_R)^\dagger$.  We start by finding an upper bound on the first term on r.h.s. above. To this end, using triangle inequality 
\begin{align}\label{eq:interim2}
\|(\mI - \mB)\mH\|_{\F} \leq \|(\mI - \mB)(\mH-\mH_R)\|_{\F}+ \|(\mI - \mB)\mH_R\|_{\F}. 
\end{align}
Using the definition in \eqref{eq:pseudo-inverse}, it is easy to verify that $\mH_R = \mH_R\mQ_1^*\big((\mQ_1\mV_R)^\dagger\big)^*\mV_R^*$. Recall from \eqref{eq:hatH} that $\mL_R$ are the top-$R$ left singular vectors of $\mY_c$ meaning that $\exists \ \mZ $ such that the best rank-$R$ approximation $\mY_{c,R}$ of $\mY_c$ is $\mY_{c,R} = \mL_R\mZ$. Its easy to check that $\mB\mY_{c,R} = \mY_{c,R}$. Using both these facts, an upper bound on the first, and the second term on the r.h.s. in \eqref{eq:interim2} are
\begin{align}\label{eq:interim3}
\|(\mI-\mB)(\mH-\mH_R)\|_{\F} &\leq \|\mI-\mB\|\|\mH-\mH_R\|_{\F}\notag\\
&\leq 3\sqrt{\frac{M}{M_2}} \|\mH-\mH_R\|_{\F},
\end{align}
and
\begin{align}\label{eq:interim4}
&\big\|\big(\mI - \mB\big)\mH_R\big\|_{\F} = \big\|\big(\mI - \mB\big)(\mH_R\mQ_1^*-\mY_{c,R})\big((\mQ_1\mV_R)^\dagger\big)^*\big\|_{\F} \notag\\
&\qquad \leq  \| \big(\mI - \mB\big)\|\|(\mH_R\mQ_1^*-\mY_{c,R})\|_{\F}\|\big((\mQ_1\mV_R)^\dagger\big)^*\|, \notag\\
& \qquad \leq 3\sqrt{2} \sqrt{\frac{M}{M_2}}\|(\mH_R\mQ_1^*-\mY_{c,R})\|_{\F},
\end{align} 
respectively, where we have used the facts that \eqref{eq:singularvalue-bounds-Q1VR},  $\|\big((\mQ_1\mV_R)^\dagger\big)^*\| \leq \sqrt{2}$, and $\|(\mI-\mB)\| \leq 1+\|(\mA_2\mL_R)^\dagger\| \leq 3\sqrt{M/M_2}$ --- using $\|\mA_2\| = 1$, and $\|\mL_R\| = 1$.  Moreover, an application of triangle inequality yields 
\begin{align}\label{eq:interim5}
&\|(\mH_R\mQ_1^*-\mY_{c,R})\|_{\F}  \leq  \|\mY_c-\mY_{c,R}\|_{\F} + \|\mY_c-\mH_R\mQ_1^*\|_{\F}\notag\\
 &\qquad\qquad \leq  \ 2 \|\mY_c-\mH_R\mQ_1^*\|\notag\\
 &\qquad\qquad \leq 2\big(\|\mY_c-\mH\mQ_1^*\|_{\F} + \|(\mH-\mH_R)\mQ_1^*\|_{\F} \big)\notag\\
&\qquad\qquad \leq 2\big(\|\mE_c\|_{\F} + \|\mQ_1\|\|\mH-\mH_R\|_{\F} \big)\notag\\
&\qquad\qquad \leq 2\Bigg(\delta_1 + \sqrt{\frac{\Delta}{\Omega}} \delta_2 + \sqrt{\frac{W}{\Omega}} \|\mH-\mH_R\|_{\F}\Bigg),
%& \big(\|(\mH-\mH_R)\mQ_1^*\|_{\F} + \|\mY_c-\mH\mQ_1^*\|_{\F}  +  \|\mY_c-\mY_{c,R}\|_{\F}\big)\\
%&\leq \big( \|\mQ_1\|\|\mH-\mH_R\|_{\F} + \|\mE_c\|_{\F} + \|\mY_c-\mY_{c,R}\|_{\F}\big)\\
%& \leq \bigg(\sqrt{\tfrac{W}{\Omega}}\|\mH-\mH_R\|_{\F} + \delta_1 + \delta_2 \sqrt{\tfrac{\Delta}{\Omega}} + \|\mY_c-\mY_{c,R}\|_{\F}\bigg),
\end{align}
where the last inequality follows from the fact that $\|\mQ_1\| \leq \sqrt{W/\Omega}$, and using \eqref{eq:Ec-bound}.  

Combining \eqref{eq:interim3}, \eqref{eq:interim4}, and \eqref{eq:interim5} with \eqref{eq:interim2}, we obtain 
\begin{align}\label{eq:first-term-bound}
\|(\mI - \mB)\mH\|_{\F} \leq %\notag\\
 6\sqrt{2}\sqrt{\frac{M}{M_2}}\Bigg(\delta_1 + \sqrt{\frac{\Delta}{\Omega}} \delta_2 + \sqrt{\frac{W}{\Omega}} \|\mH-\mH_R\|_{\F}\Bigg). 
\end{align}
As for the second term in \eqref{eq:interim1}, the upper bound is 
\begin{align}\label{eq:second-term-bound}
&\|\mK (\mY_r - \mA_2\mH)\|_{\F} \leq \|\mK\| \|\mY_r-\mA_2\mH\|_{\F},\notag\\
& \qquad \leq c \sqrt{\frac{M}{M_2}}  \left(\frac{1}{\sqrt{M_2S}}\|\mA_2\mH- (\mA_2\mH)_S\|_{1} + \delta_2\right),
\end{align}
where the second inequality is obtained by using the fact that $\|\mB\| \leq 2\sqrt{M/M_2}$, and \eqref{eq:Yr-stable-recovery}. 

Combining \eqref{eq:first-term-bound}, and \eqref{eq:second-term-bound} with \eqref{eq:interim1} completes the proof of Theorem \ref{thm:main}.

 \bibliographystyle{elsarticle-num} 
 \bibliography{CS-SC.bib}





% \end{thebibliography}
\end{document}